\documentclass[
 reprint,
 superscriptaddress,
 pdetex,
 amsmath,amssymb,
 aps,
 prl,
]{revtex4-2}
\usepackage[pdftex]{graphicx}
\usepackage[T1]{fontenc}

\usepackage{dcolumn}% Align table columns on decimal point
\usepackage{bm}% bold math
\usepackage[colorlinks=true, linkcolor=blue, citecolor=blue, urlcolor=blue]{hyperref}

\begin{document}

\preprint{APS/123-QED}

\title{Controlling particle current in a many-body quantum system by external driving}% Force line breaks with \\
%\thanks{A footnote to the article title}%

\author{Shuto Mizuno}
\affiliation{%
{\it Department of Applied Physics, Nagoya University, Nagoya 464-8603, Japan}
}%
\author{Kazuya Fujimoto}
\affiliation{%
{\it Department of physics, Tokyo Institute of Technology, Tokyo 152-8551, Japan}
}%
\author{Yuki Kawaguchi}
\affiliation{%
{\it Department of Applied Physics, Nagoya University, Nagoya 464-8603, Japan}
}%

\date{\today}

\begin{abstract}
We propose a method to control the particle current of a one-dimensional quantum system by resonating two many-body states through an external driving field.
We consider the Bose-Hubbard and spinless Fermi-Hubbard models with the Peierls phase which induces net particle currents in the many-body eigenstates. A driving field couples the ground state with one of the excited states having large net currents, enabling us to control the system's current via Rabi oscillation.
Employing the Floquet analysis, we find that the resonate excited states are determined by the symmetry of the driving field, which allows us to selectively excite only certain states among the dense spectrum of a many-body quantum system. 
\end{abstract}
\maketitle

\begin{figure}[t]
\includegraphics[width=8.6cm]{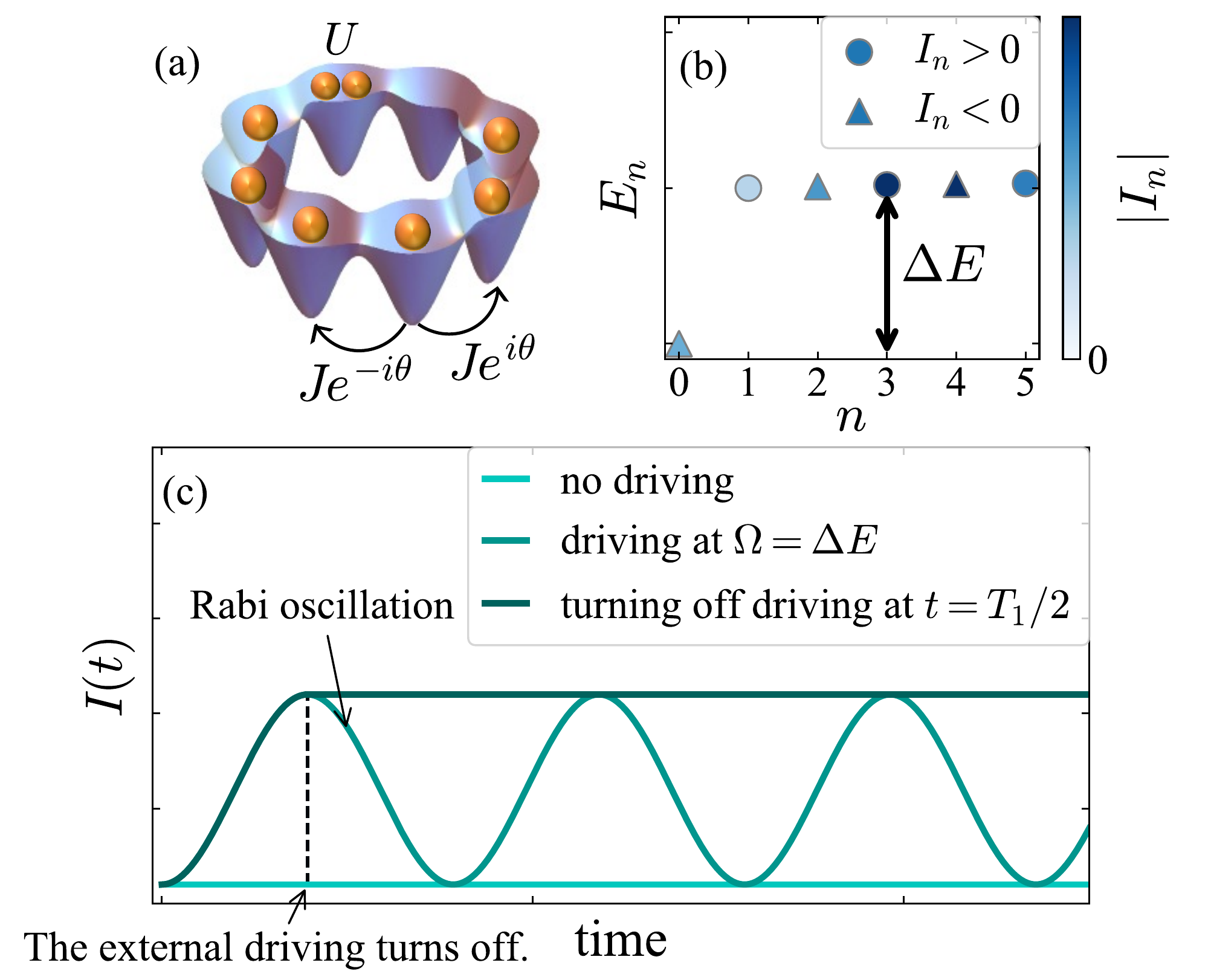}
\caption{\label{fig:schematics} 
Schematics of particle-current control for bosons. (a) Bose-Hubbard model on a ring with a Peierls phase. (b) Eigen energies at $U\gg J$ in the ascending-order $n$, where symbols $\circ$ and $\triangle$ indicate the sign of the net current in each state with color showing the amplitude. (c) Time evolution of the net particle current. The current is conserved in the absence of the driving field (light green). The driving field induces Rabi oscillation between the ground and resonant excited states (green). When we turn off the driving field at half of the Rabi oscillation period $T_1$, the system remains in the excited state (dark green).
}
\end{figure}

{\it Introduction}.---%
Developments in manipulating electric currents have been one of the central issues in modern technology. Examples range from traditional transistors to quantum-mechanical current control using Josephson effect~\cite{Josephson1962}, giant magnetoresistance effect~\cite{GMR1, GMR2}, and nonreciprocal responses~\cite{Rikken2001, Tokura2018, Ideue2021, Orenstein2021}. Recently, using oscillating fields of light has become a powerful tool to manipulate electron motion~\cite{Kirilyuk2010, Eckardt2017, Bukov2015, Oka2019, Rudner2020, Koshihara2022}. A characteristic example is the photo-induced electron transfer in molecular systems~\cite{Duncan2007, Daly2015}. It is also possible to manipulate electron transfer in superconducting materials above the transition temperature~\cite{Fausti2011, Kaiser2014, Mankowsky2014, Hu2014, Mitrano2016, Cantaluppi2018} and in Mott insulators~\cite{Okamoto2007, Wall2011, Oka2012, Mayer2015, Yamakawa2017, Murakami2018, Hejazi2019, Li2022}. 

A system of neutral cold atoms is a promising testbed for studying driving-field-induced phenomena~\cite{Eckardt2017, Weitenberg2021}. The state-of-the-art technique enables us to change the hopping integral up to its phase in an optical lattice by modulating the lattice at high frequency~\cite{Eckardt2005Analog, Eckardt2005Superfluid, Goldman2014PRX, Goldman2014RPP}, realizing the Hamiltonians of Harper-Hofstadter~\cite{Aidelsburger2013, Miyake2013} and Haldane~\cite{Jotzu2014} models. Related to the particle-current control, the phase transition from the superfluid phase to Mott insulating phase was observed using the same technique~\cite{Creffield2006, Lignier2007}. The further theoretical study discussed the current generation dynamics~\cite{Creffield2008, Sias2008}.
Though the above systems are designed so as to realize the static Hamiltonians under high-frequency driving,
one can access anomalous phases absent in static systems by lowering the driving frequency~\cite{Gao2016, Mukherjee2017, Wintersperger2020, Xie2020}. When we further lower the frequency down to the order of the system's energy, we can investigate resonance dynamics, such as parametric amplification discussed in electronic~\cite{Clerk2010, Rajasekaran2016, Okamoto2016, Babadi2017, Westig2018, Buzzi2021} and cold atomic~\cite{Engels2007, Clark2017, Lellouch2018, Gorg2018, Messer2018, Sandholzer2019, Nguyen2019, Evrard2019, Zhu2021, Kim2021} systems.

In this Letter, we theoretically propose a current-control method in a one-dimensional (1D) lattice system by resonantly transferring the system to a many-body excited state using a driving field. Differently from the previous studies of near-resonant dynamics~\cite{Engels2007, Clark2017, Lellouch2018, Gorg2018, Messer2018, Sandholzer2019, Nguyen2019, Evrard2019, Zhu2021, Kim2021}, in our method, a discrete spectrum of a finite system is essential. 
Figure~\ref{fig:schematics} schematically depicts the mechanism. We start from the ground state of bosons (spinless fermions) in a 1D ring-shaped lattice with a Peierls phase at filling fraction 1 (1/2) [Fig.~\ref{fig:schematics}(a)]. When the on-site repulsion ($U$) dominates the hopping ($J$), the ground state is a strongly-localized state with almost zero net current. There are many almost-degenerate excited states at $\Delta E\sim U$, some of which have large net currents due to the Peierls phase [Fig.~\ref{fig:schematics}(b)]. Choosing one of them and applying an oscillating field with the frequency resonant to the excitation energy, we can induce Rabi oscillation between the ground state and the excited state [Fig.~\ref{fig:schematics}(c)]. The point here is that among the numerous almost-degenerate excited states, only a small fraction of them can couple with the ground state due to the symmetry of the driving field. Although the resonant dynamics in a many-body system has been investigated for bosons in a double-well potential~\cite{Watanabe2012}, this selection rule is crucial for observing Rabi oscillation in a system with large degrees of freedom.
We also investigate the many-body dynamics at $U=0$ and find that the nonzero net current in the ground state is inverted through multiple transitions to a high-energy state.

{\it Model}.---%
We consider a system of $N$ bosons or $N$ spinless fermions in a 1D optical lattice with periodic boundary conditions, which is described with the Bose-Hubbard~\cite{Fisher1989, Jaksch1998} or spinless Fermi-Hubbard~\cite{Giamarchi2004} models:
\begin{equation}
   \label{BH}
    \hat{H}_{\text{BH}}=  -J\sum_{j=1}^{L}\left(e^{i\theta}\hat{b}_{j+1}^{\dagger} \hat{b}_{j} +\text{H.c.} \right)+\frac{U}{2}\sum_{j=1}^{L}\hat{b}_{j}^{\dagger} \hat{b}_{j}^{\dagger} \hat{b}_{j} \hat{b}_{j} ,
\end{equation}
\begin{equation}
   \label{FH}
    \hat{H}_{\text{FH}}=  -J\sum_{j=1}^{L}\left(e^{i\theta}\hat{c}_{j+1}^{\dagger}\hat{c}_{j}+\text{H.c.}\right)+U\sum_{j=1}^{L}\hat{c}_{j}^{\dagger}\hat{c}_{j}\hat{c}_{j+1}^{\dagger}\hat{c}_{j+1}.
\end{equation}
Here, $\hat{b}_{j}$ and $\hat{b}_{j}^{\dagger}$$~(\hat{c}_{j}$ and $\hat{c}_{j}^{\dagger})$ are the boson (fermion) annihilation and creation operators at site $j$, $Je^{i\theta}$ is a complex hopping  amplitude between neighboring sites with $\theta$ being the Peierls phase, and $U$ is the interaction between a pair of bosons at the same site or fermions at neighboring sites. We choose the number of lattice sites $L$ to be $L=N$ ($2N$) for the bosonic (fermionic) system such that the particles in the ground state are strongly localized for large enough $U$.

Starting from the ground state of the system, we add a time-dependent Hamiltonian given by
\begin{equation}
   \label{TD}
    \hat{H}_{1}(t)= \frac{V}{2}\sin(\Omega t)\hat{G}.
\end{equation}
Without loss of generality, we can assume $V$ and $\Omega$ to be positive. In the calculations below, we consider on-site modulation $\hat{G}=\sum_{j=1}^L K_j\hat{a}_j^\dagger\hat{a}_j$, where $\hat{a}_{j}$ denotes $\hat{b}_{j}$ ($\hat{c}_{j}$) for the bosonic (fermionic) system, and
$K_{j}$ is the site-dependent amplitude. 
The total Hamiltonian is given by $\hat{H}(t)=\hat{H}_{0}+\hat{H}_{1}(t)$, where $\hat{H}_{0}=\hat{H}_{\text{BH}}$ or $\hat{H}_{\text{FH}}$. 
Defining the net particle current operator through the system as
\begin{equation}
\label{FL}
 \hat{I}=\frac{i}{L}\sum_{j=1}^{L} {( Je^{i\theta}\hat{a}_{j+1}^{\dagger} \hat{a}_{j}-\text{H.c.})},
 \end{equation}
we discuss how its expectation value changes under the driving field. Here and in what follows, we set $\hbar=1$.

\begin{figure*}[t]
\includegraphics[width=0.95\linewidth]{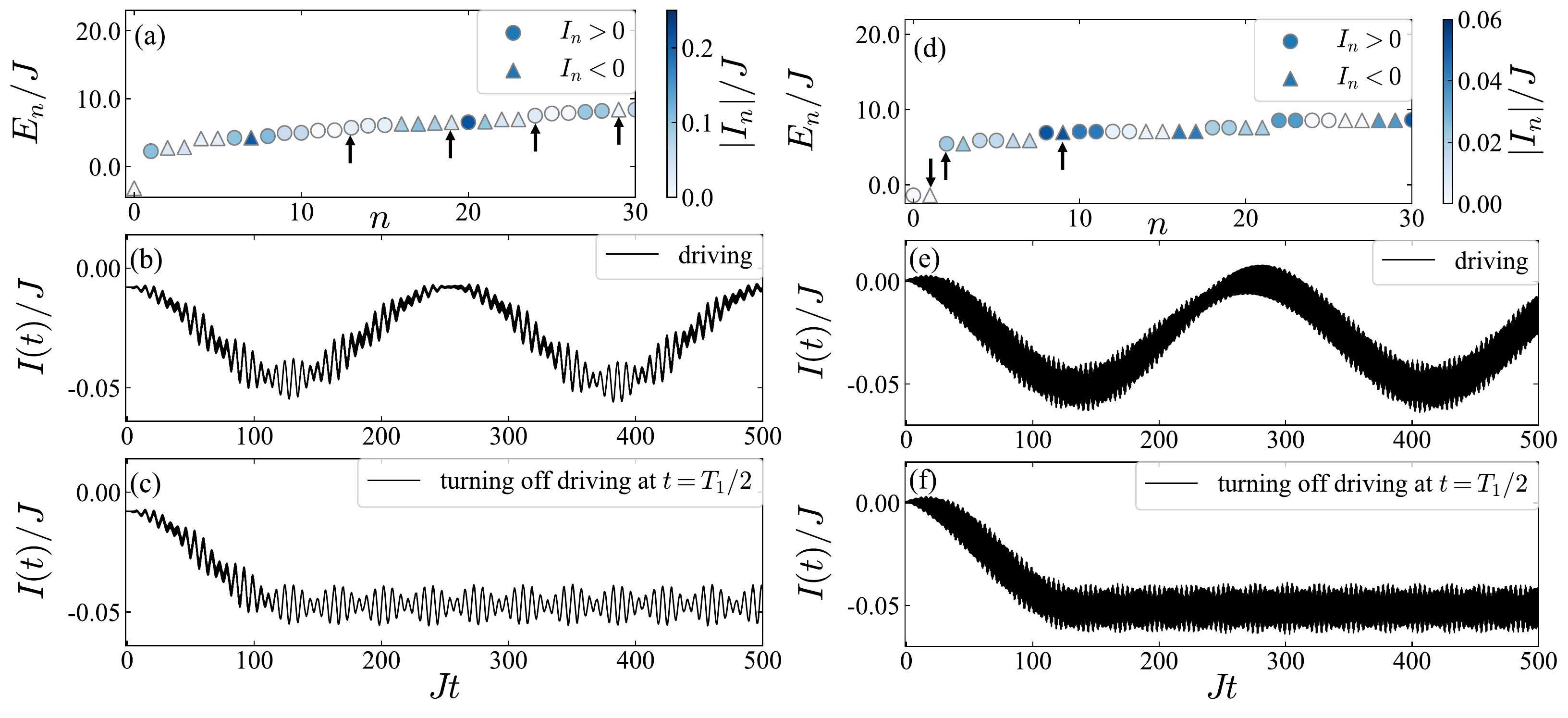}
\caption{\label{fig:Mott} Numerical results at $U/J=10$ for the Bose-Hubbard model (a)-(c) and for the spinless Fermi-Hubbard model (d)-(f). (a)(d) Eigenvalues and net particle currents of the eigenstates of $\hat{H}_0$ with $\theta=\pi/16$ for (a) and $\theta=\pi/13$ for (d). There is an energy gap between the ground and excited states. Arrows indicate excited states that resonate with the ground state under the external driving with $K_{j}=(-1)^{j}$.  (b)(e) Time-evolution of the net current after the driving field with $V/J=0.1$ is turned on at $t=0$.  The frequency of the driving field is chosen to be $\Omega=E_{19}-E_{0}$ for (b) and $E_9-E_0$ for (e). (c)(f) Same as (b) and (e) but the driving field is turned off after half of the Rabi oscillation period $T_1=4\pi/(V|{\mathcal G}_{0,n}|)$ with $n=19$ (c) and $9$ (f).
}
\end{figure*}

{\it Floquet analysis}.---%
Since the Hamiltonian is time-periodic, $\hat{H}(t)=\hat{H}(t+T)$ with period $T={2\pi}/{\Omega}$, we apply the Floquet theorem~\cite{Shirley1965, Sambe1973} to the time-dependent many-body Schr\"{o}dinger equation
\begin{equation}
   \label{Schr}
   i\frac{d}{d {t}}|\Psi(t)\rangle=\hat{H}(t)|\Psi(t)\rangle,
\end{equation}
and describe the many-body wave function as
\begin{equation}
   \label{Psi}
   |\Psi(t)\rangle = e^{-i \lambda_{F} t}\sum_{j=- \infty}^{\infty}\sum_{n=0}^{D} e^{-i\Omega j t}C_{j,n}|\phi_{n}\rangle.
\end{equation}
Here, $|\phi_n\rangle$ is the $n$th eigenstate of $\hat{H}_{0}$, i.e., $\hat{H}_{0}|\phi_{n}\rangle=E_{n}|\phi_{n}\rangle~(0\leq n \leq D$ with $D+1$ being the number of many-body eigenstates) and $\lambda_{F}$ is the quasienergy defined modulo $\Omega$. Substituting Eq.~\eqref{Psi} into Eq.~\eqref{Schr}, we obtain an infinite-dimensional eigenvalue equation for $|C\rangle =(\cdots,C_{j,0},C_{j,1},\cdots,C_{j,D},C_{j+1,0},\cdots)^{\rm T}$ as
\begin{equation}
   \label{Floquet}
  \left( \begin{array}{ccccc}
\ddots  &    \ddots                &             &                & \text{\huge{0}}  \\
 \ddots          & \mathcal{F}_{j-1}& \frac{V}{4i}\mathcal{G}&       &  \\
            &  -\frac{V}{4i}\mathcal{G}   & \mathcal{F}_{j}&  \frac{V}{4i}\mathcal{G}&       \\
            &                 &  -\frac{V}{4i}\mathcal{G} & \mathcal{F}_{j+1}&  \ddots        \\
\text{\huge{0}}&                 &                &    \ddots         & \ddots
   \end{array}\right) |C\rangle=\lambda_{F}|C\rangle,
\end{equation}
with
\begin{align}
(\mathcal{F}_j)_{m,n}&=\delta_{mn} (E_n-\Omega j),\\
\mathcal{G}_{m,n}&=\langle \phi_m|\hat{G}|\phi_n\rangle
\end{align}
being $(D+1)\times (D+1)$ matrices. 
Equation~\eqref{Floquet} means that a pair of eigenstates at $V=0$ with eigenvalues $E_m-\Omega j$ and $E_n-\Omega (j+1)$ is coupled with the coupling strength $V\mathcal{G}_{m,n}/4$ by the driving field. 
Although such terms couple an infinite number of eigenstates at $V=0$, we can approximate the system as a two-level system of $|\phi_m\rangle$ and $|\phi_n\rangle$ when $V$ is sufficiently small and $\Omega$ resonants only to the energy difference between the two states. Then, when we prepare the system in an eigenstate $|\phi_n\rangle$ and turn on the driving field, we would induce the Rabi oscillation between $|\phi_n\rangle$ and $|\phi_m\rangle$. Here, we consider starting from the ground state $|\phi_0\rangle$ and see which excited state the system can reach.

Our numerical study is two-step: First, we calculate the eigenstates of the time-independent Hamiltonian $\hat{H}_0$ by exact diagonalization. For each eigenstate, we calculate the coupling strength $\mathcal{G}_{0,n}$ with the ground state and the expectation value of the current $I_{n}=\langle\phi_{n}|\hat{I}|\phi_{n}\rangle$. Second, we numerically solve the time-dependent Schr\"{o}dinger equation~\eqref{Schr} using the Crank-Nicolson method. Using the obtained time-dependent wave function $|\Psi(t)\rangle$, we calculate the dynamics of the net particle current $I(t)=\langle\Psi(t)|\hat{I}|\Psi(t)\rangle$.

{\it Resonance at $U\gg J$}.---%
We first consider the Bose-Hubbard model with $U/J=10$, $\theta=\pi/16$, and $L=N=8$. Figure~\ref{fig:Mott}(a) shows the eigenvalue $E_{n}$ and current $I_{n}$ of each eigenstate. In the ground state, all atoms are almost localized in each site. However, since we consider a finite  system with the Peierls phase, the ground state has small nonzero current $I_0$. The currents of the most of the excited states are larger than that of the ground state. We then calculate $\mathcal{G}_{0,n}$ for the case of $K_j=(-1)^j$. The arrows in Fig.~\ref{fig:Mott}(a) indicate the excited states that have nonzero $\mathcal{G}_{0,n}$. Among them, we focus on the $n=19$th excited state. We prepare the system in the ground state $|\phi_0\rangle$ and turn on the oscillating field with $\Omega=E_{19}-E_0$ and $V/J=0.1$ at $t=0$.  The subsequent time evolution of the system's current is shown in Fig.~\ref{fig:Mott}(b). As one can see, the current oscillates between $I_0/J\cong -0.008$ and $I_{19}/J\cong -0.048$. This is due to the Rabi oscillation between the many-body eigenstates $|\phi_0\rangle$ and $|\phi_{19}\rangle$. The oscillation period at $\Omega=E_{19}-E_0$ is given by $T_1=4\pi/(V|\mathcal{G}_{0,19}|)\cong 251/J$, which agrees with the numerical result. Moreover, when we turn off the oscillating field at $t=T_1/2$, the system stays in $|\phi_{19}\rangle$ and holds the current $I_{19}$ as shown in Fig.~\ref{fig:Mott}(c). This means that we can control the particle current by transferring the system from the ground state to an excited state with a nonzero current via an oscillating field. We note that, however, the fine oscillations of $I(t)$ in Figs.~\ref{fig:Mott}(b) and \ref{fig:Mott}(c) are caused by the small mixing of the other eigenstates and are inevitable.
The mixing of other eigenstates becomes more prominent for larger $V$. In particular, when $U/J$ is large, a state at excitation energy $\sim 2U$ largely contributes to the dynamics. The detailed results are given in the Supplemental Material (SM)~\cite{SM}.

We also perform the same calculation for the spinless Fermi-Hubbard model and see the generation of the net particle current. The results corresponding to Figs.~\ref{fig:Mott}(a), \ref{fig:Mott}(b), and \ref{fig:Mott}(c) are shown in Figs. \ref{fig:Mott}(d), \ref{fig:Mott}(e), and \ref{fig:Mott}(f), respectively, where we have used $U/J=10$, $\theta=\pi/13$, $L=2N=14$, $V/J=0.1$, $K_{j}=(-1)^{j}$ and $\Omega=E_{9}-E_{0}$. 

{\it Selection rule}.---%
We note that among the numerous excited states of $\hat{H}_0$, only a small fraction of the states couple with the ground state [see arrows in Figs. 2(a) and 2(d)]. This is due to the selection rule coming from the symmetry of the driving field. The symmetry crucial for the present setup is translational symmetry.
Here, we introduce the translation operator $\hat{T}$ that moves all the particles to one site right~\cite{Sorensen2012}; It acts on $\hat{a}_{j}$ ($\hat{a}_{j}^{\dagger}$) as $\hat{T}\hat{a}_{j}=\hat{a}_{j+1}\hat{T}$ ($\hat{T}\hat{a}_{j}^{\dagger}=\hat{a}_{j+1}^{\dagger}\hat{T}$). Since $\hat{H}_0$ preserves the translational symmetry, it commutes with $\hat{T}$, i.e., the eigenstates of $\hat{H}_0$ are also the eigenstates of $\hat{T}$. We thus add a new superscript $s$ to specify the eigenvalue for $\hat{T}$ and  write the simultaneous eigenstates of $\hat{H}_0$ and $\hat{T}$ as
\begin{eqnarray}
    \label{Bloch}
    \hat{H}_{0}|\phi^{s}_{n}\rangle = E_{n}|\phi^{s}_{n}\rangle, \ \ \ 
     \hat{T}|\phi^{s}_{n}\rangle = e^{i q_{s}}|\phi^{s}_{n}\rangle,
\end{eqnarray}
where $q_{s}={2\pi s}/{L}$ ($s=0, 1, \cdots, L-1$) is the quasi-momentum of the many-body state. The discrete values of $q_s$ are obtained from the condition $\hat{T}^L=1$. Obviously from Eq.~\eqref{Bloch}, $\hat{T}$ is a unitary operator.

In Fig.~\ref{fig:Mott}, we have used $K_{j}=(-1)^{j}$, for which the operator $\hat{G}$ anti-communicates with $\hat{T}$:
\begin{equation}
 \label{Trans}
 \hat{T}\hat{G}=-\hat{G}\hat{T}.
 \end{equation}
 From Eqs.~\eqref{Bloch} and \eqref{Trans}, we obtain 
 \begin{align}
   \label{TransG}
    -\langle \phi^{s}_{0}|\hat{G}|\phi^{s'}_{n}\rangle
    =\langle \phi^{s}_{0}|\hat{T}^{-1}\hat{G}\hat{T}|\phi^{s'}_{n}\rangle
    =e^{-i (q_{s}-q_{s'})}\langle \phi^{s}_{0}|\hat{G}|\phi^{s'}_{n}\rangle,
 \end{align}
which indicates that the coupling strength $\mathcal{G}_{0,n}=\langle \phi_0^s|\hat{G}|\phi_n^{s'}\rangle$ becomes nonzero only when $e^{-i (q_{s}-q_{s'})}=-1$. Since the ground state $|\phi^{s}_{0}\rangle$ of the bosonic (fermionic) system with small $\theta$ has the quasi-momentum $q_s=0$ $(\pi)$~\footnote{For the case of the fermionic system, the ground states in the thermodynamic limit are doubly degenerate and have $q_s=0$ and $\pi$. This degeneracy is lifted in a finite system with nonzero $J$, and the $q_s=\pi$ state becomes the ground state.}, $\mathcal{G}_{0,n}$ is finite only when the excited state $|\phi^{s'}_{n}\rangle$ has the quasi-momentum $q_{s'}=\pi$ $(0)$. We numerically confirm that the excited state $|\phi_{19}\rangle$ in bosonic system ($|\phi_{9}\rangle$ in the fermionic system) considered in Fig.~\ref{fig:Mott} has $q_{s'}=\pi$ $(0)$.

We can change the quasi-momentum of the resonate excited states by changing the translational symmetry of $\hat{G}$: In the case when $\hat{T}$ and $\hat{G}$ satisfies $\hat{T}^{f}\hat{G}=\pm \hat{G}\hat{T}^{f}$ ($f\in\mathbb{N}$), the same calculation as in Eq.~\eqref{TransG} leads to $e^{-i f(q_{s}-q_{s'})}=\pm1$. For example, for the case when $L$ is an integer multiple of $2f$, $\hat{G}$ with $K_j=\text{sign}\{\sin[(j-1/2)\pi/f]\}$ satisfies $\hat{T}^f\hat{G}=-\hat{G}\hat{T}^f$. 
Using such $K_j$ with $f=2$ and 4 in the bosonic system of $L=8$ sites, we have numerically confirmed that only the excited state satisfying $e^{ifq_{s'}}=-1$ have nonzero coupling with the ground state. On the other hand, we can oscillate the interaction $U$ or the tunneling amplitude $J$ instead of the on-site potential. In such cases, $\hat{G}$ commutes with $\hat{T}$, and the system is excited to the state having the same quasi-momentum as the ground state~\cite{SM}.

We here emphasize that even though there are a number of many-body excited states, only a small fraction of them can couple with the ground state due to the symmetry property of the driving field and the excited states. In particular, consider the strongly-localized state at $U\gg J$ in an $N$-particle bosonic (fermionic) system with $L=N$ ($2N$) sites. The first band at energy gap $\sim U$ from the ground state includes $L(N-1)$ states, which are almost degenerate. We can show that the $L(N-1)$ states are divided into $L$ different quasi-momentum $q_{s'}$ states, each of which includes $N-1$ states~\cite{SM}. Thus, by controlling the energy splitting between the same $q_{s'}$ states with changing $U/J$, we can prepare a situation such that all the excited states except one are almost decoupled from the ground state under a given oscillating field.

\begin{figure}[t]
\includegraphics[width=8.6cm]{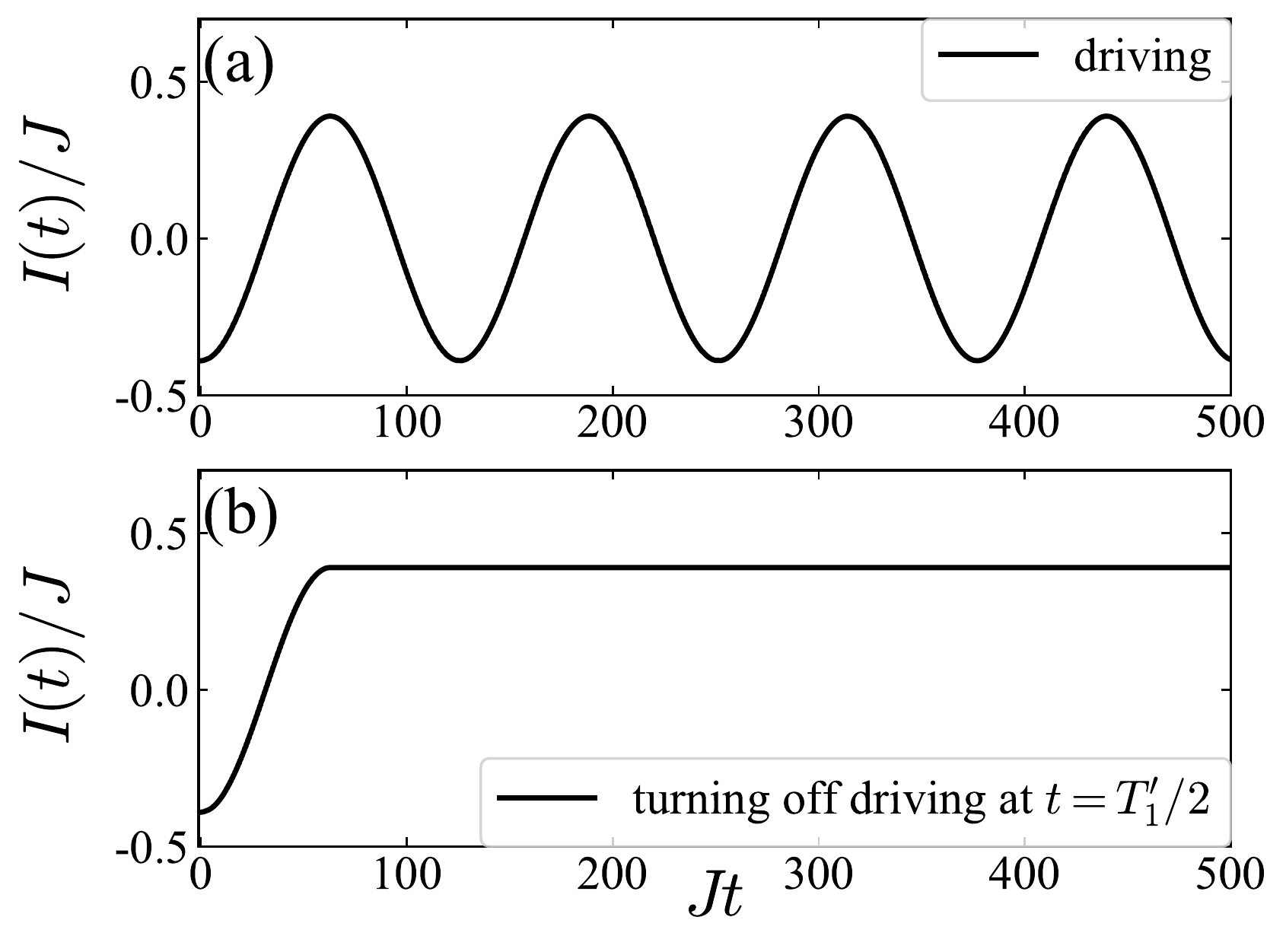}
\caption{\label{fig:U0} 
(a) Time evolution of the net particle current for the bosonic system with $U=0,\theta=\pi/16$ under a driving field of $V/J=0.1, K_{j}=(-1)^{j}$ and $\Omega=4J\cos\theta$. The initial ground state has net current $I(0)=-2J\sin\theta$. (b) Same as (a) but the driving field is turned off at half of the oscillation period $T_1'$ of $I(t)$.
}
\end{figure}

{\it Resonance at $U=0$}.---%
Finally, we consider the current-control dynamics at $U=0$ in the bosonic system. In this case, $\hat{H}_\textrm{BH}$ in Eq.~\eqref{BH} for an even $N(=L)$ reduces to
\begin{eqnarray}
   \label{FBH}
    \hat{H}_{\text{BH}} = -2J\sum^{L-1}_{s=0} \hat{b}^{\dagger}_{k_{s}} \hat{b}_{k_{s}} \cos{(\theta-k_{s})},
 \end{eqnarray}
where $\hat{b}_{k_{s}}=\sum^{L}_{j=1} \hat{a}_{j}e^{-ik_{s}j}/{\sqrt{L}}$ $(k_{s}= {2\pi s}/{L}$, $s=0, 1, \cdots, L-1)$ is the annihilation operator in momentum space. $\hat{H}_{\rm BH}$ in Eq.~\eqref{FBH} has eigenstate $\prod_{s} (\hat{b}_{k_{s}}^\dagger)^{d_{s}}/{\sqrt{d_{s}!}}|0\rangle$ with eigenenergy $-2J\sum^{L-1}_{s=0} \cos{(\theta-k_{s})}d_{s}$, where $|0\rangle$ is the vacuum state and $d_{s}(=0, 1, \cdots, N)$ represents the number of particles occupying each $k_{s}$ state and satisfies $\sum^{L-1}_{s=0}d_{s}=N$.
The ground state for $|\theta|\le \pi/L$ is given by $|\psi_0\rangle=(\hat{b}_0^\dagger)^N|0\rangle/\sqrt{N!}$.
We consider time-dependent potential~\eqref{TD} with on-site modulation $\hat{G}=\sum_{j=1}^L (-1)^{j}\hat{a}_j^\dagger\hat{a}_j$, which is  written in momentum space as 
\begin{align}
   \label{FTD}
    \hat{G}= \sum^{L-1}_{s=0} \hat{b}^{\dagger}_{k_{s}} \hat{b}_{k_ {s+{L}/{2}~\textrm{mod}~L}}.
\end{align}
Clearly, 
the oscillating field induces the components in the excited states $|\phi_{n_l}\rangle\equiv (\hat{b}_\pi^\dagger)^{l}(\hat{b}_0^\dagger)^{N-l}|0\rangle/\sqrt{l!(N-l)!}$ $(l=1, 2, \cdots, L)$ with eigenenergy $E_{n_l}=-2J(N-2l)\cos\theta$. Noting the facts that $\mathcal{G}_{n_{l-1},n_l}\neq 0$ and $E_{n_l}-E_{n_{l-1}}=4J\cos\theta$ for all $l$, the driving field with $\Omega=4J\cos\theta$ couples all the $|\phi_{n_l}\rangle$ states, and the system exhibits a periodic oscillation of the net current.
Figure~\ref{fig:U0} shows the time evolution of the net particle current for $U=0, \theta=\pi/16, L=N=8, V/J=0.1$, and $\Omega=4J\cos\theta$. 
The initial ground state has negative net current $I(t=0)=\langle\phi_0|\hat{I}|\phi_0\rangle=-2J\sin\theta$.
The driving field induces the oscillation of $I(t)$, whose largest value is $\langle \phi_{n_8}|\hat{I}|\phi_{n_8}\rangle=2J\sin\theta$ [Fig.~\ref{fig:U0}(a)]. Thus, when we turn off the driving field at half of the cycle, the net particle current is completely inverted from the ground state [Fig.~\ref{fig:U0}(b)].
We have confirmed that a small $U$ reduces the amount of the current change but keeps the qualitative behavior~\cite{SM}.

{\it Conclusions}.---%
We have studied the particle current dynamics under external driving in the 1D ring-shaped Bose-Hubbard and spinless Fermi-Hubbard models with the Peierls phase. Applying Floquet theorem to the many-body Schr\"odinger equation, 
we derived the selection rule that restricts the difference in the quasi-momenta of the coupled many-body states.
We numerically showed that a resonant driving field efficiently excites the ground state to a many-body excited state. In particular, when the ground state is strongly localized with almost zero net current at $U\gg J$, we can induce Rabi oscillation between the ground state and the resonate excited state, which can have a large net current. In the case of free bosons, the ground state has a large net current due to the Peierls phase, and the driving field can change the direction of the net current by transferring the system to a high-energy excited state via multiple resonances.

We believe that current control in the Bose-Hubbard model is experimentally feasible using a ring lattice trap~\cite{Amico2005, Amico2014}. The Peierls phase is implementable by rotating the ring lattice trap~\cite{He2009, Lacki2016}. The induced current would be measured via the time-of-flight experiment~\cite{Murray2013}. We have also examined that we can prepare the initial state by slowly introducing $J$ to the product state at $J=0$~\cite{SM}.
We remark that our method for current control is restricted to a finite system and does not apply to the thermodynamic limit with a continuous spectrum. The increase in the number of the lattice site makes the spectrum denser. Correspondingly, the number of excited states having the same quasi-momentum would increase. However, the number of resonate states for a fixed driving frequency is expected to be controllable by changing the values of $J$ and $V$.
We believe that our scheme to control the many-body wave function with external driving would be applicable for finite systems of, e.g., quantum spins, qubits, and cold atoms, including higher-dimensional systems.

\begin{acknowledgments}
This work was supported by JSPS KAKENHI (Grant Nos. JP18K03538, JP19H01824,  JP19K14628, 20H01843, 21H01009) and the Program for Fostering Researchers for the Next Generation (IAR, Nagoya University) and Building of Consortia for the Development of Human Resources in Science and Technology (MEXT). 
\end{acknowledgments}

\bibliography{main}
\bibliographystyle{apsrev4-2}

\widetext
\clearpage

\setcounter{equation}{0}
\setcounter{figure}{0}
\setcounter{section}{0}
\setcounter{table}{0}
\setcounter{page}{1}
\renewcommand{\theequation}{S-\arabic{equation}}
\renewcommand{\thefigure}{S-\arabic{figure}}
\renewcommand{\thetable}{S-\arabic{table}}

\section*{Supplemental materials for “Controlling particle current in a many-body quantum system by an external driving"}

This Supplemental Material describes the following topics:
\begin{itemize}
\item[  ]{ (I) Mixing of the other eigenstates at larger $V$, } 
\item[  ]{ (II) Current control by oscillating the on-site interaction $U$ and the tunneling amplitude $J$, }
\item[  ]{ (III) The number of many-body states having the same quasi-momentum, }
\item[  ]{ (IV) Effect of nonzero $U/J\ll 1$ on the multiple resonances, }
\item[  ]{ (V) Current dynamics starting from an experimentally realistic initial state. }
\end{itemize}

\section{Mixing of the other eigenstates at larger $V$}
We discuss the $V$ dependence of the Rabi oscillation at $U/J\gg 1$. From Eq.~\eqref{Floquet}, the strength of the driving field $V$ determines the coupling width, and the excitation to unexpected states would occur for a large $V$. This behavior is more prominent for a larger $U/J$ because the excitation spectra are more clustered at $\Delta E=U, 2U, \cdots$. 

\begin{figure}[h]
\includegraphics[width=17cm]{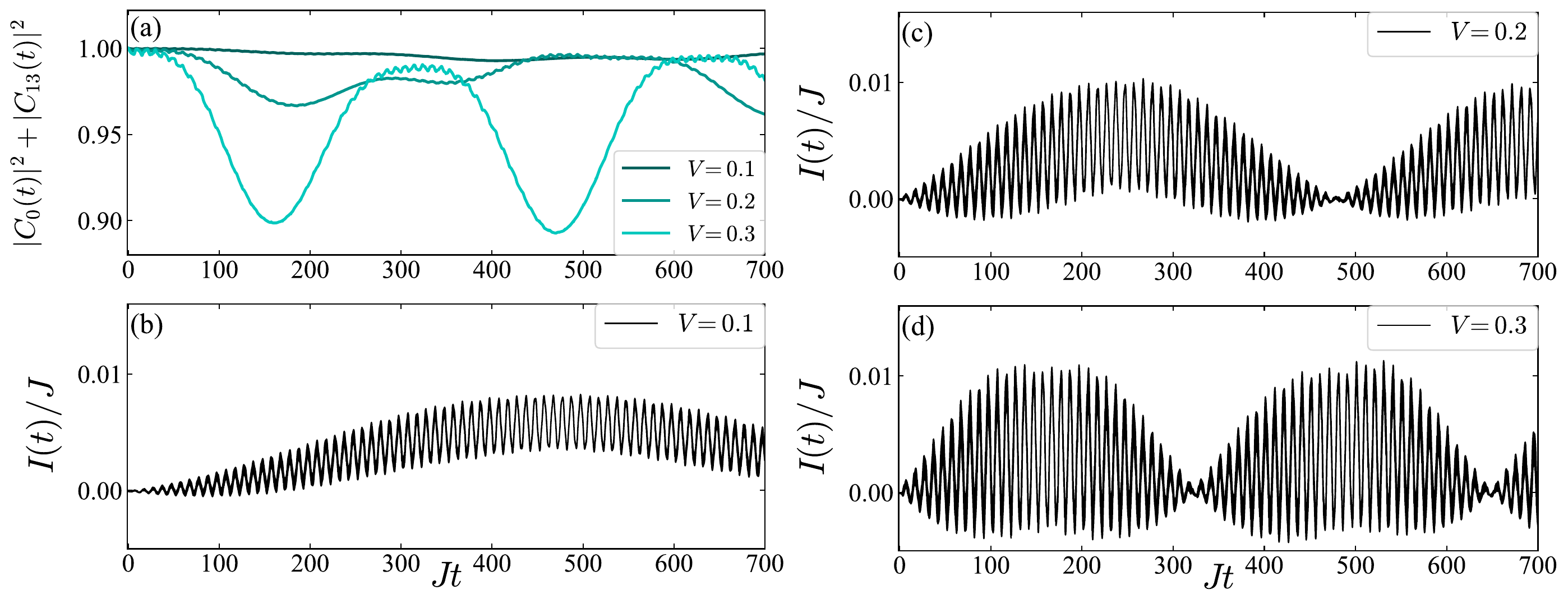}
\caption{\label{fig:U20} Numerical results for the Bose-Hubbard model with $U/J=20, \theta=\pi/16$, and $L=N=8$. (a) Time evolution of the occupancy probability in the ground state ($|\phi_{0}\rangle$) and the resonate excited state ($|\phi_{13}\rangle$) under the oscillating on-site potential with $K_j=(-1)^j$, $\Omega=E_{13}-E_0$, and
$V/J=0.1$, $0.2$, and $0.3$. (b)-(d) Time-evolution of the net particle current for the driving field with $V/J=0.1$ (b), $0.2$ (c), $0.3$ (c). The slow oscillation of $I(t)$ in (b) coming from the Rabi oscillation between $|\phi_0\rangle$ and $|\phi_{13}\rangle$ is smeared out in (c) and (d) due to the mixing of the other states.
}
\end{figure}

Here, we examine the $V$ dependence of the particle-current dynamics in the Bose-Hubbard model \eqref{BH} with $U/J=20$, $\theta=\pi/16$, and $L=N=8$. We focus on the $n=13$th excited state, which is indicated by the most-left arrow in Fig. 2(a), that has nonzero $\mathcal{G}_{0,n}$ for the case of $K_j=(-1)^j$. We prepare the system in the ground state $|\phi_0\rangle$ and turn on the oscillating field with $\Omega=E_{13}-E_0$ at $t=0$. Figure~\ref{fig:U20} (a) shows the subsequent time evolution of the occupancy probability of $|\phi_{0}\rangle$ and $|\phi_{13}\rangle$, $|C_0(t)|^2+|C_{13}(t)|^2$ with $C_n(t)\equiv \langle \phi_n|\Psi(t)\rangle$, for $V/J=0.1,0.2$, and $0.3$. Whereas the occupancy probability remains close to $1$ for $V/J=0.1$, it decreases to 0.97 for $V/J=0.2$ and to 0.90 for $V/J=0.3$. 
We have numerically confirmed that the occupancy probabilities of the 108th  and 110th excited states increase instead, which have excitation energies very close to $E_{13}+\Omega$ and have nonzero coupling strength with $|\phi_{13}\rangle$. 
For the case of $V/J=0.3$, the maximum fractions of the 108th and 110th excited states are 0.01 and 0.08, respectively.
The time evolution of the system's net current for $V/J=0.1$, $0.2$, and $0.3$ are shown in Figs.~\ref{fig:U20}(b), \ref{fig:U20}(c), and \ref{fig:U20}(d), respectively. In Fig.~\ref{fig:U20}(b), one can see the slow oscillation of the current between $I_0/J\cong -0.0001$ and $I_{13}/J\cong 0.0060$ in addition to the rapid oscillation at frequency $\Omega$. However, the oscillation of the current becomes unclear for $V/J=0.2$ and $0.3$ because of the mixing of the 108th and 110th states as shown in Figs.~\ref{fig:U20}(c) and \ref{fig:U20}(d).

\section{Current control by oscillating the on-site interaction $U$ and the tunneling amplitude $J$}
Instead of oscillating the on-site potential, we can also oscillate the on-site interaction $U$ and the tunneling amplitude $J$.
In such cases, we use 
\begin{align}
   \label{TDU}
   \hat{G}_U&=
    \sum_{j=1}^{L} \hat{b}_{j}^{\dagger} \hat{b}_{j}^{\dagger} \hat{b}_{j} \hat{b}_{j},\\
   \label{TDJ}
   \hat{G}_J&=
    -\sum_{j=1}^{L}\left(e^{i\theta}\hat{b}_{j+1}^{\dagger} \hat{b}_{j} +\text{H.c.} \right),
\end{align}
for $\hat{G}$ in the driving Hamiltonian~\eqref{TD}.
Because they commute with the translation operator $\hat{T}$,
\begin{align}
 \label{TransUJ}
 \hat{T}  \hat{G}_{U,J}=  \hat{G}_{U,J}\hat{T},
 \end{align}
 the restriction to the quasi-momentum difference between the coupled state is given by 
$e^{-i (q_s-q_{s'})}=1$.
Namely, the resonate excited states should have the same quasi-momentum as the initial state.
This resonance condition cannot be realized when we oscillate the on-site potential.

\begin{figure}[h]
\includegraphics[width=17cm]{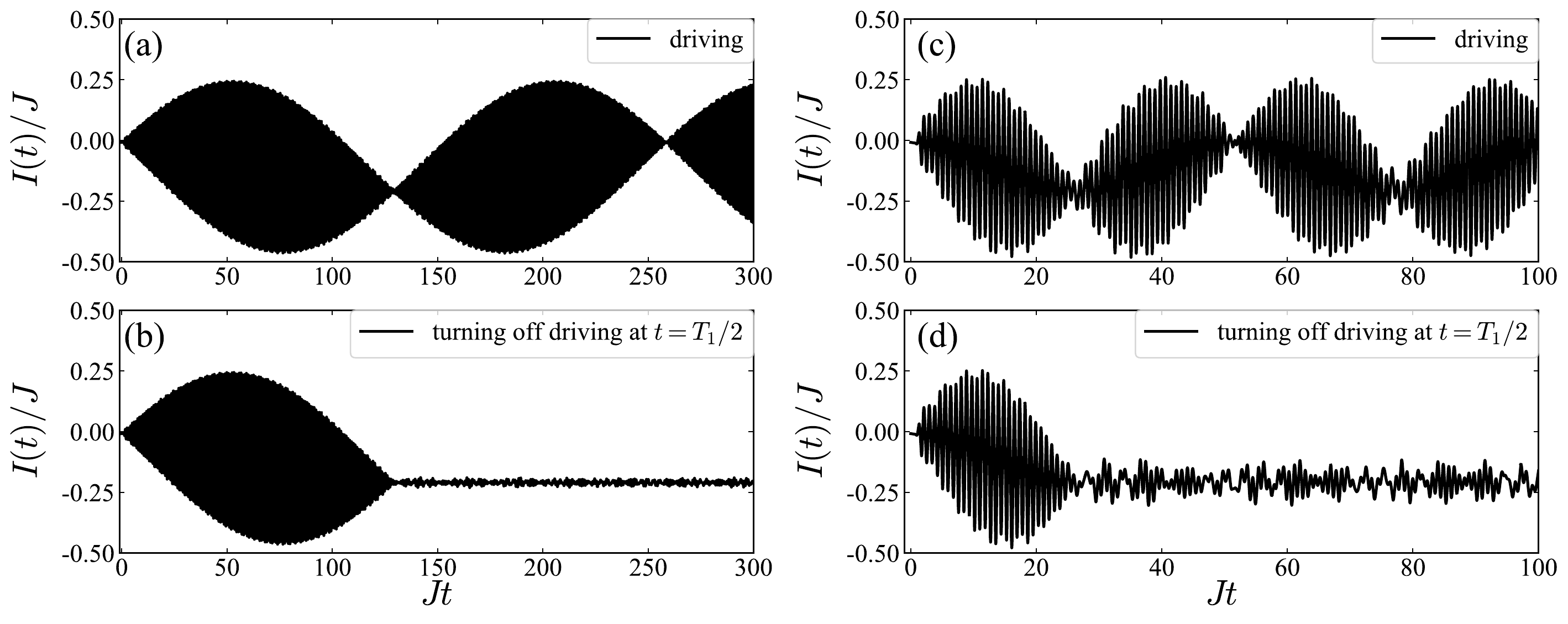}
\caption{Current dynamics under the oscillating on-site interaction $U$ (a)(b) and under the oscillating tunneling amplitude $J$ (c)(d) in the Bose-Hubbard model with $U/J=10, \theta=\pi/16, V/J=0.1$, and $\Omega=E_7-E_0$. (a)(c) Dynamics of the current under continuous driving. (b)(d) Same as (a)(c) but the driving field is turned off at half period of the Rabi oscillation $T_1$.}
\label{fig:GU_GJ}
\end{figure}

Figure~\ref{fig:GU_GJ} shows the time evolution under the driving field with $\hat{G}_U$ [Figs.~\ref{fig:GU_GJ}(a) and \ref{fig:GU_GJ}(b)] and $\hat{G}_J$ [Figs.~\ref{fig:GU_GJ}(c) and \ref{fig:GU_GJ}(d)]. The parameters in $\hat{H}_{\rm BH}$ are the same as those in Fig.~\ref{fig:Mott}(a)-\ref{fig:Mott}(c). Here, we use the resonance to the 7th excited state, which has the largest net current for this parameter setup.
Although the rapid oscillations in Figs.~\ref{fig:GU_GJ}(a) and \ref{fig:GU_GJ}(c) are amplified compared with Fig.~\ref{fig:Mott}(b),
the current is well controlled and remains constant after turning off the driving field at $t=T_1/2$.

\section{The number of many-body states having the same quasi-momentum}
We consider the case of $U/J\gg 1$ and show the numbers of the many-body excited states having the same quasi-momentum at the excitation energy $\sim U$ are the same for each quasi-momentum.

For the sake of simplicity, we first consider the case of $J=0$.
The number of states that have excitation energy $U$ is ${}_{N}C_{1}\times {}_{N-1}C_{1}=N(N-1)$ for the bosonic system with $L=N$ and $2\times {}_{N}C_{1}\times {}_{N-1}C_{1}=2N(N-1)$ for the fermionic system with $L=2N$.
At $J=0$, all the Fock states are the eigenstates of the Hamiltonian $\hat{H}_0$.
The ground-state energy is zero,
with one boson per site or one fermion per two sites.
The lowest excitation at energy $U$ is a doublon-holon excitation, where one of the particles is excited to the other occupied (vacant) site in the bosonic (fermionic) system, which is also described with a Fock state.
Let us denote one of such excited Fock states by $|n\rangle$.
Then, we immediately obtain $L-1$ different excited Fock states belonging to the eigenvalue $U$ as $\hat{T}^k|n\rangle$ ($k=1, 2, \cdots, L-1$), where $\hat{T}$ is the translation operator introduced in the main text.
We consider a unitary transformation of these $L$ states,
  \begin{equation}
  \label{eq:SM_q}
    |\phi^s\rangle=\sum_{k=0}^{L-1}C_{sk}\hat{T}^{k}|n\rangle,\ \ (s=0, 1, \cdots, L-1)
\end{equation}
and require that the new eigenstate $|\phi^s\rangle$ is the eigenstate of $\hat{T}$ with eigenvalue $e^{i2\pi s/L}$.
Such a unitary transformation is given by
\begin{align}
\label{eq:SM_q2}
    C_{sk}=\frac{1}{\sqrt{L}}e^{-i2\pi sk/L}.
\end{align}
Indeed, we can show
\begin{align}
    (C C^\dagger)_{ss'}= \sum_{k=0}^{L-1} C_{sk} (C_{s'k})^* = \frac{1}{L} \sum_{k=0}^{L-1} e^{-i2\pi(s-s')k/L}=\delta_{s,s'}
\end{align}
and
 \begin{align}
    \hat{T}|\phi^s\rangle
    &=\frac{1}{\sqrt{L}}\sum_{k=0}^{L-1}e^{-i2\pi sk/L}\hat{T}^{k+1}|n\rangle\nonumber\\
    &=\frac{e^{i2\pi s/L}}{\sqrt{L}}\sum_{k=0}^{L-1}e^{-i2\pi s(k+1)/L}\hat{T}^{k+1}|n\rangle  \nonumber\\
    &=\frac{e^{i2\pi s/L}}{\sqrt{L}}\sum_{k=0}^{L-1}e^{-i2\pi sk/L}\hat{T}^{k}|n\rangle  \nonumber\\
    &=e^{i2\pi s/L}|\phi^s\rangle,
\end{align}
where we have used $\hat{T}^L=1$ in the third equality.
The above procedure means that from an eigenstate of $\hat{H}_0$, $|n\rangle$, which breaks the translational symmetry,
we can construct $L$ eigenstates of $\hat{H}_0$ that are the eigenstates of $\hat{T}$ with $L$ different eigenvalues.
Considering the fact that the $L(N-1)$ eigenstates belong to energy eigenvalue $U$, 
there are $N-1$ states corresponding to $|n\rangle$ in the above argument, from each of which we can construct $L$ different quasi-momentum state.
Thus, among the $L(N-1)$ eigenstates at energy $U$, the number of the states that have the same quasi-momentum $q_s=2\pi s/L$ is $N-1$, which is the number of states that can couple with the ground state under a given driving field.

We can expand the above argument for the case of $J\neq 0$ and $U/J\gg 1$. The particles in the ground state are still strongly localized at each site.
However, because of the existence of the nonzero hopping $J$, the Fock state $|n\rangle$ in Eq.~\eqref{eq:SM_q} is no more the eigenstate of $\hat{H}_0$.
The hopping Hamiltonian preserves the quasi-momentum of the many-body eigenstates, which means that
in the leading order of $J/U$, the eigenstate of $\hat{H}_0$ should be given by a linear combination of states having the same quasi-momentum, where each term is described as Eq.~\eqref{eq:SM_q}.
Although there exist small contributions from the higher-energy excitations, they do not change the quasi-momenta of the many-body eigenstates at excitation energy $\sim U$.
Hence, the number of the eigenstates having the same quasi-momentum at energy $\sim U$ is the same as that for $J=0$.

\section{Effect of nonzero $U/J\ll 1$ on the multiple resonances}
\begin{figure}[h]
\includegraphics[width=17cm]{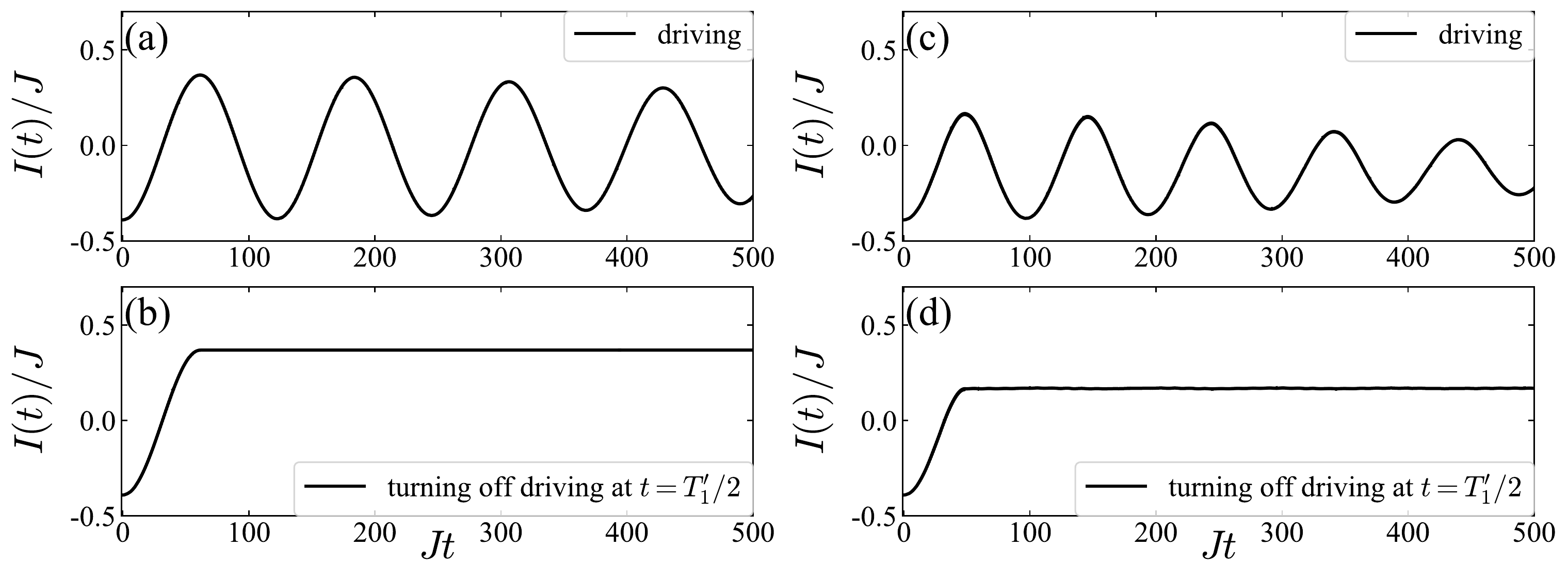}
\caption{\label{fig:U/Jll1}(a)(c)Time evolution of the net particle current for the bosonic system with $U/J \ll 1$ and $\theta=\pi/16$ under a driving field of $K_{j}=(-1)^{j}$ and $V/J=0.1$. The interaction $U$ and the frequency of the driving field are chosen as $U/J=0.01, \Omega=E_{63}-E_{0}$ for (a) and $U/J=0.05, \Omega=E_{62}-E_{0}$ for (c). (b)(d) Same as (a) and (c) but the driving field is turned off at half of the oscillation period of $I(t)$.}
\end{figure}

We have shown in the main text that for the case of $U=0$, the net particle current in the initial ground state will be converted by the driving field via multiple resonances. Here, we discuss the effect of a nonzero $U$. In Fig.~\ref{fig:U/Jll1}, we show the numerical results for $U/J=0.01$ [Figs.~\ref{fig:U/Jll1}(a) and \ref{fig:U/Jll1}(b)] and $0.05$ [Figs.~\ref{fig:U/Jll1}(c) and \ref{fig:U/Jll1}(d)].
As  $U/J$ increases, the amplitude of the oscillation of $J(t)$ becomes smaller. However, the direction of the net current is still converted.

\section{Current dynamics starting from an experimentally realistic initial state}
\begin{figure}[h]
\includegraphics[width=17cm]{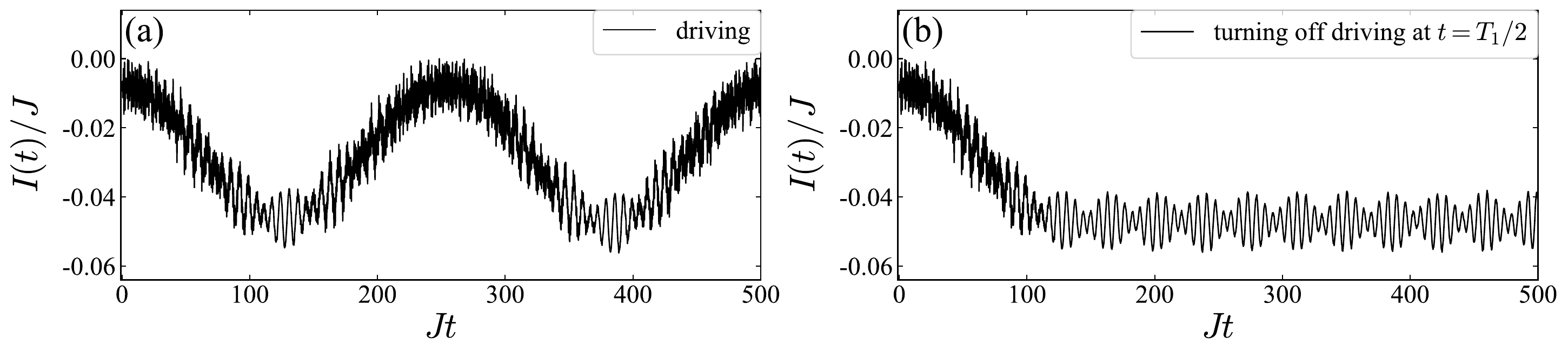}
\caption{
\label{fig:initialstate}
Time evolution of the net particle current for the bosonic system starting from the experimentally realistic initial state.
We obtain the initial state by preparing
the product state of the one-particle state in each site at $J=0$ and ramping up the value of $J$ from 0 to 1 as $J(t)=0.1t$. 
The other parameters are the same as those in Figs.~\ref{fig:Mott}(a) and \ref{fig:Mott}(b), i.e., we use
$U=10,\theta=\pi/16,K_{j}=(-1)^{j}, V=0.1$, and $\Omega=E_{19}-E_{0}$.
(a)Current dynamics under continuous driving. (b)Current dynamics when the driving field is turned off at half period of the Rabi oscillation.}
\end{figure}

In the main text, we have used the exact ground state of the system as the initial state for the time evolution. 
However, it is difficult to prepare such initial state experimentally.
To see how the initial state preparation affects the subsequent dynamics,
we start from the product state of the one-particle state in each site at $J=0$ and ramp up the value of $J$.
Here, we choose $U=10$, $\theta=\pi/16$, and change the value of $J$ from $0$ to $1$ as $J(t)=0.1t$. 
Figure~\ref{fig:initialstate} shows the subsequent dynamics of the net particle current after turning on the same driving field as that for Figs.~\ref{fig:Mott}(b) and \ref{fig:Mott}(c).
Compared with Figs.~\ref{fig:Mott}(b) and \ref{fig:Mott}(c),
the amplitude of the rapid oscillation with frequency $\Omega$ increases
due to the small mixing of the other many-body eigenstate in the initial state.
However,
our main result still holds; Rabi oscillation between the many-body eigenstates leads to the oscillation of the net current and we can control the current by the driving field. 

\end{document}